\documentclass[12pt]{article}
\usepackage{amsmath,amssymb}
\usepackage{amsthm}
\usepackage{cite}
\usepackage{amsfonts}
\def\CD{{\cal D}}

\def\CF{{\cal F}}

\def\CO{{\cal O}}

\def\CV{{\cal V}}

\def\CZ{{\cal Z}}

\newcommand{\bbibitem}[1]{\bibitem{#1}\marginpar{#1}}

\def\Label#1{\label{#1}%
  \smash{\hbox to0pt{\raise1ex\hbox{\tiny[#1]}\hss}}}
\def\noLabels{\let\Label=\label}
\def\nobbibitem{\let\bbibitem=\bibitem}

\newcommand{\be}{\begin{equation}}
\newcommand{\ee}{\end{equation}}
\newcommand{\bea}{\begin{eqnarray}}
\newcommand{\eea}{\end{eqnarray}}
\newcommand{\pat}{\partial}

\newcommand{\zbar}{\bar{z}}

\newcommand{\Nbar}{\bar{N}}
\newcommand{\Ebar}{\bar{E}}
\newcommand{\patb}{\bar{\partial}}
\def\half{\frac{1}{2}}
\def\p{\partial}

\def\bra{\langle}
\def\ket{\rangle}

\def\D{{\rm D}}

\newcommand{\refb}[1]{(\ref{#1})}

\begin{document}
\noLabels
\nobbibitem

%\begin{flushleft}
%\today
%Dec. 10, 2006
%\end{flushleft}

%title
\rightline{hep-th/0612090} \rightline{HIP-2006-53/TH} \rightline{UPR-1169-T}
\vskip 1cm \centerline{\Large {\bf A Thermodynamic Interpretation
of Time}} \centerline{\Large {\bf for Rolling Tachyons}} \vskip
0.5cm
\renewcommand{\thefootnote}{\fnsymbol{footnote}}
\centerline{{\bf Vijay
Balasubramanian,$^{1}$\footnote{vijay@physics.upenn.edu}
Niko Jokela,$^{2}$\footnote{niko.jokela@helsinki.fi}}}
\centerline{{\bf Esko Keski-Vakkuri,$^{2,3}$\footnote{esko.keski-vakkuri@helsinki.fi} and
Jaydeep Majumder$^{2}$\footnote{jaydeep.majumder@helsinki.fi}
}}
\vskip .5cm
\centerline{\it {}$^{1}$David Rittenhouse Laboratories, University of
Pennsylvania}
\centerline{\it Philadelphia, PA 19104, U.S.A.}
\vskip .5cm
\centerline{\it ${}^{2}$Helsinki Institute of Physics and ${}^{3}$Department of Physical Sciences
} \centerline{\it P.O.Box 64, FIN-00014 University of Helsinki, Finland}

\setcounter{footnote}{0}
\renewcommand{\thefootnote}{\arabic{footnote}}

\begin{abstract}
We show that the open string worldsheet description of brane decay (discussing a specific example of a rolling tachyon background) can be related to a sequence of points of thermodynamic equilibrium of a grand canonical ensemble of point charges on a circle, the Dyson gas. Subsequent instants of time are related to neighboring values of the chemical potential or the average particle number $\Nbar$.
The free energy of the system decreases in the direction of larger $\Nbar$ or later times,
thus defining a thermodynamic arrow of time. Time evolution equations are mapped to differential
equations relating thermal expectation values of certain observables at different points of thermal
equilibrium.   This suggests some lessons concerning emergence of time from an underlying microscopic structure in which the concept of time is absent.
\end{abstract}

\newpage
\tableofcontents

\section{Introduction}

There are a number of examples in string theory in which several
dimensions of space emerge as an effective description of a system
that does not contain them
\cite{c=1review,matrixreview,adscftreview}. All these  examples
arise at their root from the duality between open and closed
strings, by equivalently describing the dynamics of a system of
D-branes in terms of the open string degrees of freedom that
quantize them, or in terms of the closed strings that quantize the
spacetime that the D-branes create. There are no similar examples of
the emergence of time\footnote{For recent discussions of emergence
of spacetime and string theory, see
\cite{Seiberg:2006wf,Polyakov:2006bz}. For a recent discussion from
a condensed matter physics perspective, see \cite{Gu:2006vw}.}. A
candidate setting for investigating such an emergence is provided by
the unstable branes of bosonic and superstring theory
\cite{senrolling}. In this case, open strings are localized in {\em
time} \cite{Gutperle:2002ai}, and one might ask how the open strings
contrive to describe closed strings and spacetime at times far from
the region of time where they are confined.

Previous work showed that the worldsheet correlation functions of
strings in such universes can be computed at tree level from an
ensemble of $U(N)$ matrices
\cite{nlt,Okuyama:2003jk,Gutperle:2003xf,bkkn,Shelton:2004ij,Jokela:2005ha}.
As the brane decays to a gas of closed strings, the rank of the
matrices contributing to the ensemble increases to infinity. In
examples of the emergence of space
\cite{c=1review,matrixreview,adscftreview}, closed string
backgrounds emerge from the dynamics of $U(N)$ matrices in the large
$N$ limit. This suggests that for decaying branes the geometry at
late times is ``emerging'' from the dynamics of the large rank
matrices appearing in the open string description.

Here we take a related perspective, by showing that all worldsheet correlation functions in decaying brane universes can also be computed at tree level in string theory from a simple statistical mechanical
model -- test charges interacting with a Dyson gas, consisting of point charges confined to a circle.
Physical time maps into (an analytic continuation of) the fugacity in the gas.    Thus, by a Legendre transform, time becomes related to the mean number of particles ($\bar{N}$) in the statistical system.   (In turn, this is related to the rank of the matrices appearing in the matrix model description of the system \cite{nlt,bkkn}.) The free energy of the Dyson gas decreases with $\bar{N}$, giving rise to a thermodynamic arrow for time.

In classical physics, we are used to specifying the state of a system by specifying its initial conditions
and then allowing the equation of motion to evolve the configuration through time.
Since the Dyson gas arises here from a rewriting of worldsheet perturbation theory, the spacetime
equations of motion can be derived  by requiring conformal invariance (perhaps implying criticality) of the statistical system.   The time evolution equations of spacetime fields then translate into equations relating the values of certain statistical observables at different points of thermodynamic equilibrium.

As a rewriting of {\it worldsheet} perturbation theory in a different, statistical language,
our analysis does not describe a holographic duality in the sense of the AdS/CFT correspondence.
The latter is a relationship between the {\it spacetime} theory of open strings ({\em i.e.} a quantum
field theory) and a {\it spacetime} theory of closed strings.  In principle, we are simply describing a dictionary between two equivalent {\it worldsheet} formulations.   Nevertheless, the appearance of a thermodynamic arrow of time from our statistical formulation of the decaying brane is suggestive of some kind of emergent phenomenon.    The discussion in Sec.~4, attempts to draw lessons about what it could mean for time to be ``emergent''.

\section{The half S-brane as a Dyson Gas}

\subsection{The half S-brane as an ensemble of matrix models}

Bosonic string theory has a spectrum of unstable D-branes that decay by tachyon condensation.  Sen \cite{senrolling} showed that their homogeneous decay can be described in terms of the open string theory on a D-brane by the exactly marginal boundary deformation
\be
 \delta S_{{\rm bdry}} = \lambda \int dt~e^{X^0(t)} \ .
 \Label{boundarydef}
\ee
The parameter $\lambda$ has no meaning since it can be absorbed by redefining
the origin of the target space time coordinate $X^0$.    The resulting background is called the half S-brane since it represents a spacelike D-brane at $t \to -\infty$ that decays into closed strings as time passes.

Worldsheet correlation functions in the background of the decaying brane take the form
\begin{equation}
\bar{A}_l = \int DX^0 \, DX^1 \cdots DX^{25} \, e^{-S} \, \prod_{a=1}^l V_a(z_a,\bar{z}_a)
\end{equation}
where the action $S$ includes the boundary deformation (\ref{boundarydef}) and the $V_a$ are vertex operators constructed from $\{ X^0, \vec{X} \}$ and written as functions of holomorphic coordinates on the worldsheet.   String scattering amplitudes are obtained by integrating these correlation functions over all vertex operator positions (after fixing the location of one operator by using the conformal symmetry of the worldsheet).

Taking $V_a$ to be tachyon vertex operators and leaving out the trivial spatial part of the
computation, the $X^0$ dependent piece of the correlator is
\begin{equation}
\bar{A}_l \sim \int DX^0 \, e^{-S} \, \prod_{a=1}^l e^{i k^0_a X^0(z_a, \bar{z}_a)}\,.
\Label{tachyonamp}
\end{equation}
At non-zero momentum
it is possible to choose a gauge in which general  on-shell closed string vertex operators with
finite energy have the form
\begin{equation}
V = e^{i k^0 X^0} V_{{\rm sp}},
\Label{noX0gauge}
\end{equation}
where $V_{{\rm sp}}$ is constructed entirely out of the 25 spatial fields \cite{gauge}.
In this gauge, and given that the boundary perturbation  only depends on $X^0$, the non-trivial part of
finite momentum string scattering amplitudes can be written as (\ref{tachyonamp}).

The worldsheet correlation function (\ref{tachyonamp}) can be evaluated on the disk by isolating the
zero mode $x^0$ from the fluctuations as $X^0 = x^0 + X^{\prime 0}$, and expanding the boundary
perturbation $e^{-\delta S_{{\rm bdry}}}$ in a power series. The result is
\begin{equation}
\bar{A}_l = \int dx^0 e^{ix^0 \sum_a k^0_a} \sum_{N=0}^\infty \frac{(-2\pi \lambda e^{x^0})^N}{N!}
\int \prod_{i=1}^N \frac{dt_i}{2\pi}
\langle e^{X^{\prime 0}(t_1)} \cdots e^{X^{\prime 0}(t_N)}
\prod_{a=1}^l e^{i\omega_a X^{\prime0}(z_a,\bar{z}_a)} \rangle.
\Label{tachyonamp2}
\end{equation}
Now using the standard  Neumann correlator of $X'^0$ (see, e.g.,
Sec.~2.2 and 3.1 of \cite{bkkn}) the disk expectation value in
(\ref{tachyonamp2}) can be evaluated as
\begin{eqnarray}
\bar{A}_l &=&
 \int dx^0 e^{ix^0 \sum_a \omega_a}  \, A_l(x^0) \\
A_l(x^0) &=&
\prod_{a<b} |z_a - z_b|^{-k^0_a k^0_b}
\prod_{ab} |1 - z_a \bar{z}_b|^{-\half k^0_a k^0_b}
 F(z_1,k^0_1; \cdots ; z_l,k^0_l;x^0),
\Label{tachyonamp3}
\end{eqnarray}
where the initial factors involving $z_a$ arise from contractions between the vertex operators,
the integral over $x^0$ is the Fourier transformation to total energy, and $F$ isolates all the interesting pieces
in the correlator arising from the interaction between the vertex operator and the boundary
perturbation (\ref{boundarydef}).   $A_l(x^0)$ is the amplitude written as a function of target space time.
%Now defining
%\begin{equation}
%e^{2\mu'} = -2\pi\lambda e^{x^0}
%\end{equation}
We find
\begin{eqnarray}
F(\{z_a,k^0_a\}; x^0)
&=& \sum_{N=0}^\infty \frac{(-2\pi\lambda e^{x^0})^N}{N!}
\int \prod_{i=1}^N \frac{dt_i}{2\pi}  \prod_{i<j} |e^{it_i} - e^{it_j}|^2
\prod_{ia} |1 - z_a e^{-it_i}|^{2ik^0_a}
%\Label{F1}
\nonumber \\
&=&
\sum_{N=0}^\infty \frac{(-2\pi\lambda e^{x^0})^N}{N!}
\int_{U(N)} dU \prod_a |\det(1 - z_a U)|^{2ik^0_a} \, .
\Label{F2}
\end{eqnarray}
The second equality arises by recognizing
the measure for integration over $U(N)$.

In this way,  worldsheet correlation functions for a decaying brane are expressed in terms of an
ensemble of $U(N)$ matrices with varying rank $N$.
%a chemical potential $\mu$ that decreases with target space time $x^0$
%\todo{This is no longer true with the definition of $\mu'$ we have given.}.
As a simple example, the disk partition function is
\begin{eqnarray}
Z_{open} &=&  \bra e^{-\delta S_{{\rm bdry}}} \ket = \sum_{N=0}^\infty \frac{(-2\pi\lambda e^{x^0})^N}{N!}
\int_{U(N)} dU
\Label{Zopen1} \\
&=& \frac{1}{1 + 2\pi \lambda e^{x^0}}  \, .\Label{diskpf}
\end{eqnarray}
At early times $x^0 \to -\infty$  the first few terms in the series (\ref{Zopen1}) dominate and
hence only low rank $U(N)$ matrices contribute.   However, as $x^0$ increases, and brane decay
progresses further, matrices of larger rank become progressively more important.  The series
converges when
$|2\pi \lambda e^{x^0}|<1$. However, the summed expression (\ref{diskpf}) can be analytically
continued to future infinity $x^0\rightarrow \infty$.
The convergence radius $x^0 \sim -\ln (2\pi \lambda )$ might be interpreted as the moment
after which the tachyon has completely condensed, so that the open string worldsheet is no longer
really meaningful.  After this point, a closed string description is better.  Indeed, the function
\begin{equation}
f(x^0)= \frac{1}{1+2\pi\lambda e^{x^0}}
\end{equation}
is the coefficient of the lowest,
closed string tachyon component of the  boundary state of
the decaying brane \cite{senrolling, nlt}\footnote{For additional discussion and
early references, see \cite{Gutperle:2003xf,Callan:1994ub,Gaberdiel:2001xm}.}.

As another example of an interesting target space observable,
consider the energy-momentum tensor as a function of target space
time $x^0$ \cite{Chen:2002fp,nlt}. It can be decomposed as
\be
  T^{MN}(x^0) = K [\eta^{MN}{\cal B}(x^0) + {\cal A}^{MN} (x^0) ] \ ,
  \Label{tmunu}
\ee
where $ {\cal B}(x^0) = f(x^0)$ is the open string disk partition function, whereas
\be
 {\cal A}^{MN}(x^0) = 2\bra :\partial X^{M} (0)\bar\partial X^{N} (0): \ket_{\half S-brane}
 \label{amunu}
\ee
is the disk one-point function of the closed string vertex operator
\be
  V^{MN}(z,\zbar;k) = :\partial X^M (z) \bar\partial X^{N} (\zbar) e^{ik\cdot
  X(z,\zbar)}:
\ee
at zero momentum $k=0$, placed at the origin $(z,\zbar)=0$.  (Since this is at zero momentum we
cannot write this in the gauge (\ref{noX0gauge}).)   The perturbative expansion is again expressed
in terms of an ensemble of $U(N)$ matrices with a finite convergence radius as before \cite{nlt,bkkn}.
The series summation gives, for example,
the ${\cal A}^{00}$ component
\be
{\cal A}^{00} (x^0) = f(x^0) -2 \ ,
\ee
which can then be analytically continued to late times.

In general, all closed string observables can be computed using the open string worldsheet theory,
using similar perturbative expansions in terms of an ensemble of matrix models with a finite convergence
radius, and the sum can be analytically continued to late times.     For example, one-point functions
of closed string  on-shell vertex operators correspond to probabilities to produce closed string states
in the decay \cite{Lambert:2003zr}. The functions obtained can also be interpreted as coefficients of the
associated component in the decaying brane boundary state \cite{Gutperle:2002ai,Okuda:2002yd}.
The resulting  closed string
description is then valid at all times, even if the open string formalism is strictly valid only within
the convergence radius, at suitably early times.

\subsection{The half S-brane as a Dyson Gas: partition function}

We can now show that all the S-brane disk amplitudes can also be expressed in terms of the correlation functions in a simple statistical mechanical system --   the classical Coulomb gas of infinitely heavy unit charges on a unit circle, also called the {\em Dyson gas}.   In particular, the S-brane disk partition function is related to the grand canonical partition function of the Dyson gas (see Appendix D of \cite{bkkn} for a brief discussion).

To see this, recall first that a Dyson gas \cite{Dyson:1962es}
consists of heavy unit charges confined to live on a
unit circle in a two-dimensional plane.    Pairs of charges, having positions $e^{it_i}$, interact via the two-dimensional  Coulomb potential which is logarithmic.   The kinetic term is zero since the particles are taken to be infinitely heavy and they do not move.   The repulsive interactions give a two-body potential
\be
  V(t_i,t_j) = -\ln |e^{it_i}-e^{it_j}| \ .
  \Label{2body}
\ee
The grand canonical partition function of the Dyson gas is
\begin{eqnarray}
Z_G &=& \sum_{N=0}^\infty \frac{z^N}{N!}  Z_N(\beta)
\Label{ZG1}
\\
&=& \sum_{N=0}^\infty \frac{z^N}{N!}
\int_0^{2\pi} \prod_{i=1}^N \frac{dt_i}{2\pi} \, \exp(-\beta \sum_{k<l}^N V(t_k,t_l)),
\Label{ZG2}
\end{eqnarray}
where $z$ is the fugacity and $Z_N(\beta)$ is the canonical partition function
for $N$ particles as a function of the inverse temperature $\beta = 1/T$.
Putting in the logarithmic potential (\ref{2body}), we see that
at the special temperature $\beta^{-1} = 1/2$,
\begin{eqnarray}
Z_G &=& \sum_{N=0}^\infty \frac{z^N}{N!}
\int \prod_{i=1}^N \frac{dt_i}{2\pi} \left[ \prod_{i<j} |e^{it_i} - e^{it_j}|^2 \right]
\Label{ZG3}
  \\
&=& \sum_{N=0}^\infty \frac{z^N}{N!}  \int_{U(N)} dU = \frac{1}{1 -z},
\Label{ZG4}
\end{eqnarray}
where we used the identification of the integral over $U(N)$ as in (\ref{F2}) and (\ref{Zopen1}).
Thus we see that for $\beta = 2$, the interactions between the Dyson gas particles reproduce
the integrations over $U(N)$ in (\ref{F2}).

Comparing (\ref{diskpf}) and (\ref{ZG4}) it is tempting to identify $z= -2\pi \lambda e^{x_0}$ and $Z_G = Z_{open}$.
However, this is inconvenient because fugacity is the exponential of a chemical potential,
and hence this identification would require an imaginary component for the chemical potential in our Dyson gas.   Rather, $Z_G$ is more naturally identified with another quantity in the decaying brane worldsheet theory, namely  the expectation value of an operator $(-1)^{{\hat N}_T}$, where ${\hat N}_T$ counts the number of tachyon vertex operators from the rolling tachyon background:
\bea\label{notdiskpf}
 Z'_{open} = \bra (-1)^{{\hat N}_T} \ket_{\half S-brane} &=& \sum^\infty_{N=0} (-1)^N(-2\pi \lambda e^{x^0} )^N \nonumber \\
                        &=& \frac{1}{1-2\pi \lambda e^{x^0}} \ .
\eea
Identifying $Z_G \leftrightarrow Z'_{open}$ gives
\be
  z = e^{\beta \mu} = 2\pi \lambda e^{x^0} \ ,
  \Label{ztime}
\ee
so the chemical potential $\mu$ essentially corresponds to target space time $x^0$. Thus, considering the decaying brane at different times corresponds to considering the Dyson gas at different points of thermodynamic equilibrium,   corresponding to different values of the chemical potential in the grand canonical ensemble,

The interesting physical quantities in the target space of the decaying brane are computed as expectation values of closed string vertex operators
with respect to the disk partition function $Z_{open}$ rather than $Z'_{open}$.    In order to obtain these observables from the Dyson gas formulation, we will need to compute the grand canonical partition function $Z_G$ as a function of a conventional fugacity and then analytically continue
\be
   z \rightarrow ze^{i\delta} \rightarrow ze^{i\pi} = -z
\Label{contin}
\ee
to negative values of $z$, while continuing to identify time $x^0$ via the absolute value of $z$ as in (\ref{ztime}).    Recall that because the late time physics ($x^0 \to \infty$) lies beyond the convergence radii of the series in (\ref{Zopen1}) and (\ref{ZG4}), an analytic continuation in the complex fugacity plane was in any case required.   The continuation (\ref{contin}) relating the string partition function $Z_{open}$ to the Dyson gas partition function $Z_G$ is simply an extension of this procedure.

\subsection{The half S-brane as a Dyson Gas: correlation functions}

Above we showed how the open string disk partition function is obtained from the thermodynamics of the Dyson gas.  A general closed string correlation function is obtained by inserting vertex operators in the bulk of the disk.
For homogenous brane decay, as described by (\ref{boundarydef}),
the spatial fields $X^i,\ i = 1, \ldots, 25$ in the vertex operator will simply contract amongst themselves and give a standard result on the disk.   Thus, as in (\ref{tachyonamp}), we will leave out this trivial spatial part of the computation since we are interested in the dependence of correlators on target space time.   Then, the relevant part of the general vertex operator can be written as sums of products of terms of the general form
\begin{equation}
:\partial^{n_1} X^0(z) \bar{\partial}^{n_2} X^0(\bar{z}) e^{i k_0 X^0}:.
\end{equation}
The correlation functions of such operators can be computed in the Dyson gas approach by inserting
external charges at the locations of the vertex operators and computing the expectation values of
suitable moments of the potential.

To illustrate, consider the stress tensor ($n_1=n_2=1$ and $k^0\rightarrow 0$).   The $T^{00}$ component
contains  the disk partition function $Z_{open}={\cal B}(x^0)$ and the expectation
value ${\cal A}^{00}= 2\bra :\partial X^0 \bar \partial X^0 :\ket$ (\ref{tmunu}).  In the Dyson gas
thermodynamics, the latter quantity is computed by placing an additional  charge  $k^0$ at
$(z_0,\bar{z_0})$  on the plane, inside the circle. The charged particles of the Dyson gas on the unit circle will produce an
electrostatic potential acting on the bulk charge,
\be
 V^{{\rm bulk-Dyson}} = -k^0\sum_k \ln |z_0-e^{it_k}|^2 \ .
\ee
The holomorphic and anti-holomorphic derivatives of $V^{{\rm bulk-Dyson}}$ will be called
holomorphic and antiholomorphic ``forces",
\be
  F^{{\rm bulk-Dyson}} = \frac{\pat V^{{\rm bulk-Dyson}}}{\pat z_0} \ , \
  \bar{F}^{{\rm bulk-Dyson}} = \frac{\pat V^{{\rm bulk-Dyson}}}{\pat \zbar_0} \ .
\ee
To compute ${\cal A}^{00}$ in the Dyson gas, compute the grand canonical expectation value of the
product of these ``forces'',
\bea
 \bra\bra F(z_0) \bar{F}(\zbar_0) \ket\ket_\beta & \equiv & \sum^{\infty}_{N=0} \frac{z^N}{N!} \int \prod_{i=1}^N
\frac{dt_i}{2\pi} \exp [ -\beta (V^{{\rm Dyson}} +  V^{{\rm bulk-Dyson}}) ]\nonumber\\
 & & \cdot F^{{\rm bulk-Dyson}}(z_0,\{ t_i\})\bar{F}^{{\rm bulk-Dyson}}(\bar{z}_0,\{ t_i\}) \ .\Label{thermexp}
\eea
then move the external charge to the origin, $z_0=\zbar_0=0$, and continue in the complex fugacity plane
\be
 z\rightarrow -z = -2\pi \lambda e^{x^0} \ ,
\ee
and replace\footnote{Charges must be continued to imaginary values in order to make contact with
real energies.} $k^0 \to ik^0$ to finally obtain
\be
   \bra\bra F(z_0) F(\zbar_0) \ket\ket_{\beta}
   \longrightarrow \bra :\pat X^0 \patb X^0 :\ket = \half \cdot {\cal A}^{00}(x^0) \ .
\ee
This is a prescription for computing the target space stress tensor for a decaying brane in terms of
expectation values in the grand canonical ensemble for a Dyson gas.
Here
\be
V^{{\rm Dyson}} = -\sum_{k<l} \ln | e^{it_k } - e^{it_l}|
\ee
is the potential energy of the Dyson gas.

It is straightforward to generalize this to a comparison of more
general thermal expectation values and more complicated closed and
open string correlation functions.  We add external charges at the
locations of the vertex operators and compute the grand canonical
expectation value of an appropriate moment of $V$.    The potential
between the external charges will produce an overall  worldsheet
position dependence in the amplitude ({\em e.g.}, the initial
factors in (\ref{tachyonamp3})), while the interactions with the
Dyson gas particles produce the part of the amplitude that is
affected by the decay of the brane ({\em e.g.}, the final factor $F$
in (\ref{tachyonamp3})). Finally, to compute a string amplitude, the
positions of all but one of the vertex operators are integrated over
the disk.\footnote{We need 3 real parameters to fix the residual
$SL(2,\mathbb{R})$ symmetry.} This treatment can be applied to open
string vertex operators too, except that here the external charges
are inserted into the circle on which the Dyson gas particles are
already present. The prescription is illustrated by some examples in
the table below.

%%%%%%    Table %%%%%%%%
\renewcommand{\arraystretch}{1.0}
\begin{table}[!ht]
\begin{center}
\begin{tabular}{||c|c||}\hline
In Rolling Tachyon CFT & In Dyson Gas Ensemble \\\hline\hline
                           & Insert external particle with charge $p^0$ on the \\
$:e^{ip^0X^0(\tau)}:$      & circle and modify the potential term by adding an  \\
(on the boundary)          & interaction term for this external \\
                           & charged particle and Dyson gas \\\hline
                                      & Insert charged particle on the\\
$: \p^n X^0(\tau) :$        & boundary and compute the thermal average of\\
                                      & the $n$-th moment of the potential in grand \\
                                      & canonical partition function of the Dyson gas\\\hline
                            & Insert particle with charge $k^0$ on 2d plane\\
$:e^{ik^0X^0(z,\bar{z})}:$  & inside the circle at $(z,\bar{z})$ and modify the potential \\
(in the bulk)               & by adding interaction between \\
                            & the bulk charge and Dyson gas particles\\\hline
                            & Insert external charged particle \\
$ :\partial^{n_1}X^0(z)\bar{\partial}^{n_2}X^0(\bar{z}):$
                            & inside the circle and compute thermal average of \\
                            & $n_1$-th holomorphic and $n_2$-th anti-holomorphic \\
                            & moments of the potential in the grand \\
                            & canonical partition function of the Dyson gas\\\hline
\end{tabular}
\end{center}
\caption{{\sl Correspondence between state in CFT and thermal
average in Dyson gas. Left hand side of the table is to be
evaluated inside a correlator in rolling tachyon background. All
the vertex operators in bosonic string theory can be obtained by
taking suitable combinations of the above set.}}
\end{table}

\section{The flow of time}

In the Dyson gas formulation, target space time has been replaced by
the chemical potential $\mu$ of a statistical system.    Different
points of thermodynamic equilibrium characterized by different
values of $\mu$ (keeping the temperature and the volume fixed)
correspond to the different instants in time for the decaying brane.

\subsection{Time as the average particle number}

It is now simple to give an interpretation of time as a measure of
the average particle number in the Dyson gas.   In terms of the
fugacity $z$, the average particle number is:
 \begin{equation}
  \Nbar = z\partial_z \ln Z_G = \frac{z}{1-z} \ .
  \Label{Nz}
 \end{equation}
Recall now that every positive fugacity value was in one-to-one
correspondence with an instant of time $x^0$ for the decaying brane,
by
\be
  |z| = 2\pi\lambda e^{x^0} \ .
\ee
So the average number of particles in the Dyson gas related to the instant $x^0$ is
given by
 \begin{equation}
  \Nbar = \frac{2\pi \lambda e^{x^0}}{1- 2\pi \lambda e^{x^0} } \  \  \
    \Longrightarrow \  \  \
  x^0 =  \ln\left(\frac{\bar{N}}{1 + \bar{N}}\right) - \ln 2\pi\lambda \ .
  \end{equation}
In particular, the infinite past $x^0=-\infty$ corresponds to vanishing fugacity and  $\Nbar$.
The average particle number then increases monotonically as a function of the fugacity,
corresponding to later time values $x^0$. This relation is valid up to the point $z=1$ where
$\Nbar$ diverges. This corresponds to time value $x^0 = - \ln 2\pi\lambda$. But we have also
seen that this is also the boundary of the convergence
radius for typical open string worldsheet calculations, beyond which it is more appropriate to
use the closed string description of the system.

The average
particle number $\Nbar$ is a continuous quantity, just like time.
Interestingly, however, the underlying physical quantity in the
Dyson gas is the actual number of particles $N$.  This is {\it
discrete}, and fluctuates around $\Nbar$. The relative size of the
fluctuations declines as $\Nbar$ increases:
\begin{equation}
  \delta N = \frac{\sqrt{\overline{N^2}-\Nbar^2}}{\Nbar}
           = \frac{1}{\sqrt{z}} = \frac{e^{-x^0/2}}{\sqrt{2\pi \lambda}} \ .
\end{equation}
At early times (small $\Nbar$),  the relative fluctuations of $N$
are large and at later times $\Nbar$ becomes more sharply defined.
It is tempting to interpret this as ``a continuous time emerging
from an underlying discrete variable in the large $N$ limit".\footnote{Quantitatively, one can cut off the infinite series summations at some value $N_{max} \gg \Nbar$ and get good approximations to all calculations.     The size of the cutoff  $N_{max}$ increases with time.}

\subsection{A thermodynamic arrow of time}

Consider the grand potential of the Dyson gas ensemble, \be
  \Omega(\mu,T,V) = -T \ln Z_G(\mu,T,V) \ ,
\ee where we set Boltzmann's constant to 1. $\Omega$ satisfies the
thermodynamic identity \be
 \Omega = \Ebar -TS -\mu \Nbar \ ,
\ee where $\Ebar$ is the mean energy of the system. A Legendre
transformation gives the conventional Helmholtz free energy:
\begin{equation}
 A(\Nbar,T,V) = \Omega + \mu \Nbar = \bar{E} - TS \  ,
\end{equation}
where we substitute for the chemical potential $\mu$ as a function
of $T,\Nbar$ by inverting (\ref{Nz}). $A$ measures the free energy
of the system as a function of the particle number (as opposed to
the chemical potential).   Setting $T=1/2$ for the Dyson gas, we
find that \be\label{helmholtz}
 A(\Nbar,T,V) = -\frac{1}{2} \left[(\Nbar+1)\ln (\Nbar+1)-\Nbar\ln\Nbar\right].
\ee
This resembles  the familiar formula for entropy as a function
of the mean number of particles.   Indeed, the Shannon entropy in the particle number distribution is
\begin{equation}
I_S = -\sum_n p(n) \ln p(n) ~~~~~;~~~~~ p(n) = \frac{z^n}{Z_G} \ .
\Label{shannon}
\end{equation}
where $p(n)$ is the probability that there are $n$ charged particles in the Dyson gas.
Doing the sum (\ref{shannon}) and using (\ref{Nz}) for the mean particle number, we find
\begin{equation}
I_S = (\Nbar+1)\ln (\Nbar+1)-\Nbar\ln\Nbar ~~~~~ \Longrightarrow ~~~~~~
A(\Nbar,T,V) = - \, T  \, I_S \ .
\end{equation}
The Shannon entropy increases with  $\Nbar$ and hence the free energy decreases with the mean
particle number.    Recall that the increase of $\Nbar$ marks the passage of time.  Thus the free
energy decreases with the passage of time.  We interpret this as a thermodynamic arrow of time for
decaying brane universes.

\subsection{Equations of motion: generalities}

Usually,  the flow of time is discussed in terms of a {\it time evolution} equation that determines
the configuration of a system at all times, given an initial condition.  Hence,  it is interesting
to ask how the spacetime equations of motion appear in the Dyson gas formulation of decaying branes.

In string theory the spacetime fields appear as coupling constants in the worldsheet sigma model,
and their equation of motion arises by requiring vanishing
of all the beta functions \cite{Alvarez-Gaume:1981hn}.
In the presence of an unstable brane, the worldsheet theory includes a tachyonic boundary deformation.
It is more subtle to derive the beta function in the presence of such a source,
there are additional corrections
which can be viewed to arise from higher order string diagrams in a degenerate limit \cite{Callan:1986bc,
Leigh:1989jq}.
The vanishing of the corresponding beta function then leads to a solution of the spacetime equations of motion
containing a decaying brane.  Specifically, the exactly marginal boundary
deformation (\ref{boundarydef}) preserves worldsheet conformal invariance
and leads to a specific time dependent source
term in the spacetime equations of motion.  For example, the dilaton
field equation in the presence of the decaying brane is
\be
  \nabla^2 \phi = J_{\rm{dil}}
    =\bra V_{\phi}(0) e^{-\delta S_{{\rm bdry}}} \ket_{\rm{disk}} \ ,
\ee
where the r.h.s. is the one-point function of the dilaton vertex
operator in the presence of the boundary deformation.
For the homogenously rolling tachyon and a decaying $p$-brane in
bosonic theory, the field equation turns out to be \cite{Maloney:2003ck}
\be
  \eta^{MN}\partial_{M}\partial_{N}\phi (x^0,\vec{x})
  = c\cdot \delta^{25-p}(\vec{x})f(x^0) \ ,
\ee
where $M,N=0,\ldots,25$, $c$ is a numerical constant and $f$ is related to the
target space stress tensor. Focusing on the
simplest case of a space-filling brane $(p=25)$, the equation
simplifies to \be
  -\partial^2_0 \phi (x^0) = c\cdot f(x^0) \ .
  \Label{dileqn}
\ee

Since we have essentially reformulated worldsheet
perturbation theory in terms of a statistical mechanical system,
it should be possible to interpret equations like (\ref{dileqn})
in terms of Dyson gas thermodynamics.   Recalling that the time
$x^0$ is related to the chemical potential as $x^0 \sim 2\mu$, and
that $f$ is interpreted in terms  of a thermodynamic average of
moments of the electrostatic potential between the gas charges and
an external charge at the origin, we expect an equation in the
Dyson gas of the form
\be
  -\partial^2_{\mu} \phi (\mu) = 4c\cdot f(\mu) \ .
  \Label{dileqntherm}
\ee
In other words, if we identify the quantity $\phi$ as an object in
the Dyson gas, its value at all points of equilibrium (fixed by
different values of $\mu$) can be determined via (\ref{dileqntherm})
in terms of an ``initial value'' $\phi(\mu_0)$ and the initial
gradient $\partial_\mu\phi(\mu_0)$.

The identification of the spacetime fields within the Dyson gas
formulation and the derivation of the equations of motion is done
below and in the Appendix by translating the worldsheet beta
function formulation into our statistical language.   We have been
unable to find a clear account in the literature of the derivation
of the beta function equations for bulk fields in the presence of
a source created by a tachyonic boundary deformation.  While a complete
calculation would clearly be an interesting basic string theory problem,
the details are tangential to the discussion here.  We outline the steps of the
calculation in the Appendix.

\subsection{Equations of motion: derivation}

We begin by recalling how the field equations arise from requiring
Weyl invariance of the decaying brane worldsheet sigma model.
We start with the worldsheet action
\bea\label{action}
  S &=& S_P + \delta S_{{\rm bdry}} \nonumber \\
  \mbox{} &=& \frac{1}{4\pi\alpha'} \int d^2\sigma \sqrt{h} h^{ab} \eta_{MN}
  \partial_a X^M \partial_b X^N + \lambda \int ds~e^{X^0} \ .
\eea
Next, we compute the disk partition function, as in (\ref{diskpf}),
but now the contractions between operators at the boundary will
involve the curved space generalization $d_h(s_i,s_j)$
of the flat space distance function
$d(e^{it_i},e^{it_j})=|e^{it_i}-e^{it_j}|$. The correlators
between bosons on the boundary become
\be
\bra X^0 (s_i) X^0 (s_j) \ket = \Delta^{{\rm bdry}}_h (s_i,s_j)
=  \ln d^2_h(s_i,s_j) \ .
\ee
The disk partition function then becomes
\bea
  Z_{{\rm{disk}}} = Z_{\rm{sp}}\times \sum^{\infty}_{N=0} \frac{(-2\pi\lambda e^{x^0})^N}{N!}
  \int \prod_i\frac{ds_i}{2\pi}
 \prod_{i<j} \exp \left(  \Delta^{{\rm bdry}}_h(s_i,s_j) \right) \ ,
\eea where $Z_{\rm{sp}}$ is the trivial contribution from the 25
free spacelike bosons $X^i$.

The corresponding statistical mechanical system (in addition to the
free boson CFT) is the curved space generalization of the Dyson
gas, a system where unit point charges on a (locally scaled) circle
interact through a two-body Coulomb potential in curved space,
\be
  V^{{\rm Dyson}}_h
  = -\half \sum_{i<j}
 \Delta^{{\rm bdry}}_h(s_i,s_j) \ ,
\ee
with the canonical partition function
\be
 Z_N(\beta ) = \int\prod^N_{i=1} \frac{ds_i}{2\pi} \exp (-\beta V^{{\rm Dyson}}_h )
   \ .
\ee
The inverse temperature is again $\beta=2$.  In the curved case
there is no obvious random matrix interpretation.  As before, we
must consider a grand canonical ensemble partition function to make
contact with the worldsheet theory.

The next step in deriving field equations is to introduce background
fields in the worldsheet sigma model, for example a general target
space metric $G_{MN}(X)$ and a dilaton $\phi (X)$, \be S[G,\Phi] =
\frac{1}{4\pi\alpha'}\int d^2\sigma \, \sqrt h \, [h^{ab} \,
G_{MN}(X) \, \partial_a X^M\partial_b X^N + \alpha'  \,
\widetilde{R} \, \Phi(X)]\,, \ee where $\widetilde{R}$ is the
worldsheet Ricci scalar. We take the target space metric to be
almost flat, \be \label{metric}
  G_{MN}(X) = \eta_{MN} + \epsilon_{MN} (X) \
\ee
and choose $\epsilon_{MN}(X)$ to be a monochromatic gravitational plane wave
\be
   \epsilon_{MN}(X) = -4\pi \, g_c \, \varepsilon_{MN} \, e^{ik\cdot X} \ ,
\ee
where $g_c$ denotes the string coupling.
Similarly, we take a plane wave of the dilaton:
\be
\Phi(X) = -4\pi \,  g_c \, \phi \, e^{ik\cdot X} \ .
\ee
Next, perform a Taylor expansion in the metric and dilaton perturbation for the disk partition function, so that
it becomes
\be
\label{taylorexp}
  Z_{\rm{disk}}[\varepsilon,\phi] =  \langle e^{-\delta S_{bdry}} \, \{ 1 + g_c~:\CV:
 + \frac{g_c^2}{2!}~:\CV\CV:
   + \cdots \} \rangle_{\eta}
\ee
where
\be
\label{vk}
  \CV = \frac{g_c}{\alpha'}
  \int d^2 \sigma\sqrt{h(\sigma)}\left[\varepsilon_{MN} \,  h^{ab}(\sigma ) \,
 \partial_a X^M \partial_b X^N \, e^{ik\cdot X} + \alpha' \, \phi \, \widetilde{R} \,
e^{ik\cdot X}\right] \ee is  the graviton-dilaton vertex operator.
Requiring $Z_{{\rm disk}}$ to be invariant under the Weyl
transformations $h_{ab} \to h'_{ab} = e^{2\omega(\sigma)}h_{ab}$
then imposes the on-shell conditions and physical polarization
conditions for $\varepsilon_{MN}$. Or, more precisely, in coordinate
space the $\beta$-function equation gives the (linearized) Einstein
equation (see {\em e.g.} \cite{Polchinski:1998rq}) in the presence
of the rolling tachyon source {\em in string frame}:\footnote{The
derivation of the source term is subtle.  See the Appendix.} \be
\label{einstein} R_{MN}^L + 2\nabla_M^L\nabla_N^L \phi = 8\pi
T_{MN}^L \ . \ee Here the superscript $L$ means linearized. Recall
that the gravitational field from which the Ricci tensor is
constructed is (\ref{metric}); to first order in $\epsilon$ \be
\label{ricciweak} R_{MN}^L = \half \left\{ \pat^2 \epsilon_{MN}
-\pat_P \pat_M \epsilon^P_{\ N} -\pat_P \pat_N \epsilon^P_{\ M}
 + \pat_M\pat_N \epsilon^P_{\ P} \right\} \ .
\ee
Meanwhile, the stress tensor $T_{MN}$ appearing on the right
side of (\ref{einstein}) is equal to the disk one-point functions
(\ref{tmunu})-(\ref{amunu}).

We now have sufficient information to identify what computation in
the Dyson gas gives the thermodynamic counterpart of the Einstein
equation (\ref{einstein}). Generating a weakly curved background
spacetime metric corresponds to introducing additional bulk charges
to the system, and considering the thermal expectation values of
some (generalized) forces resulting from the point charges on the
circle acting on the added test charge.  Then, just like above, we
should require scale invariance to derive the counterpart of the
field equations. It is possible that the requirement of scale
invariance is tantamount to requiring that the statistical system be
at a critical point, but we have not explored this carefully.    In any case,
deformations of the Dyson gas which introduce additional external charges result in deformations of the spacetime.

In order to keep the exposition brief, we will ignore the 25 free
spacelike bosons that play role in determining the full Einstein
equation (\ref{einstein}).
Thus, in what follows we will only focus
on the time-time component
\be
\label{einstein00}
R_{00} + 2\partial_0^2\phi = 8\pi T_{00} \ ,
\ee
as it will be straightforward to repeat the
calculations for the other components.

Our starting point is (\ref{taylorexp}), which instructs us to
consider the quantity \bea \label{cg}
Z_{{\textrm{``disk''}}}[\varepsilon,\phi] \equiv
Z_{sp}\times\left(\bra\bra 1 \ket\ket_\beta
  + g_c \bra\bra{\cal O} \ket\ket_\beta + \cdots \right) \, ,
\eea where $Z_{\rm{sp}}$ denotes the contribution from the 25
spatial directions of (\ref{taylorexp}), and the rest is the
non-trivial contribution from the grand canonical ensemble of
charges. We can calculate that the second term in the expansion,
corresponding to the second term $\bra e^{-\delta S_{{\rm bdry}}}
\int dk~\CV(k)\ket$ in (\ref{taylorexp}), is \bea
\bra\bra\widehat{{\cal O}} \ket\ket_\beta \equiv  z^{k^0}
\sum^{\infty}_{N=0} \frac{z^N}{N!}\bra\bra\widehat{{\cal O}}_N(k^0)
\ket\ket_{\beta ,N} \ , \eea where \bea\label{thermalavg} &&
\bra\bra\widehat{{\cal O}}_N(k^0) \ket\ket_{\beta ,N} \equiv
\frac{1}{2}\varepsilon_{00} \int d^2\sigma \sqrt{h(\sigma )} \int
\prod^N_{i=1} \frac{ds_i}{2\pi}
\exp \Big\{-\beta \Big[:V_h^{{\rm bulk}}(\sigma,\sigma'):|_{\sigma'\to\sigma}+ \nonumber \\
&& \ \ \ \ \ \ \ \ \ \ \ \ \
\ \ \ \ \ \  V^{{\rm Dyson}}_h +  V^{{\rm bulk-Dyson}}_h \Big]\Big\}\nonumber\\
&\times& \Bigg[\ h^{ab}: \partial_a V^{{\rm bulk}}_h:|_{\sigma'\to\sigma}
:\partial_{b} V^{{\rm bulk}}_h:|_{\sigma'\to\sigma}
  + 2\ \frac{1}{k^0}\ \epsilon^{ab}\
\partial_a V^{{\rm bulk}}_h\ \partial_{b} V^{{\rm bulk-Dyson}}_h
 \nonumber \\
&& \ \ \ \ \ \ \ \ \ \ \ \ \ \ \   + \frac{1}{(k^0)^2}
h^{ab}\partial_a V^{{\rm bulk-Dyson}}_h\
 \partial_{b} V^{{\rm bulk-Dyson}}_h
- \frac{1}{2k^0}\ h^{ab}:\partial_a\partial_{b}V^{{\rm bulk}}_h:|_{\sigma'\to\sigma}
  \Bigg]\ . \nonumber
\eea
In the above, $\sigma = (\sigma^1,\sigma^2)$ is the position of the bulk charge inside the disk,
$\epsilon^{ab}$ is the unique antisymmetric tensor in two dimensions
 and
$$
:V_h^{{\rm bulk}}(\sigma,\sigma'):|_{\sigma'\to\sigma} +  V^{{\rm bulk-Dyson}}_h
$$
is a regularized\footnote{We subtract the usual divergent selfinteraction contribution of the
bulk charge
but keep the finite contribution from its image charge necessitated by the Neumann
boundary condition.} net potential energy measured at $\sigma$ due to a bulk charge $k^0$
placed at $\sigma$ and Dyson gas on the boundary of the unit disk, subject to the Neumann
boundary condition on the boundary.
%\todo{VB: What does regularized mean here?}
These terms can be expressed in terms of curved space generalization of
bulk-bulk $\bra X^0(\sigma)X^0(\sigma')\ket$
and bulk-boundary $\bra X^0(\sigma)X^0(s_i)\ket$ potentials respectively:
\bea
V^{{\rm bulk-Dyson}}_h &=& -k^0\sum_i\Delta^{{\rm bulk-bdry}}_h(\sigma,s_i)\nonumber\\
V_h^{{\rm bulk}}(\sigma,\sigma') &=& -k^0\Delta^{{\rm
bulk}}_h(\sigma,\sigma')\,, \eea where both $\Delta^{{\rm
bulk-bdry}}_h(\sigma,s_i)$ and $\Delta^{{\rm
bulk}}_h(\sigma,\sigma')$ are subject to the Neumann condition on
the boundary. The expression inside the square bracket of
(\ref{thermalavg}) is the product of the generalized forces on the
bulk charge from the Dyson gas. Those expressions are regularized
since these are measured at the location of the bulk charge itself.
We denote this regularization scheme by $:\ :$.   Finally, to get
the expression of (\ref{taylorexp}) from (\ref{thermalavg}), we
follow these two rules\footnote{Note also that the spectator CFT
contained in $Z_{{\rm sp}}$ is required to fully recover the answer
of (\ref{taylorexp}).}: (1) Analytically continue $z \to -z =-
2\pi\lambda e^{x^0}$,  (2) Replace $k^0$ by $ik^0$.

In the thermodynamical system,  the above quantities are thermal
expectation values of particular moments of the Coulomb interaction
potentials.  Requiring (\ref{cg}) to be scale invariant yields the
Einstein equation (\ref{einstein}). (For additional details that will be
involved in the derivation, see the Appendix.)  Derived in
this way, the Einstein equation does not refer to spacetime
tensors; rather it refers to Fourier transformations of quantities
in the momentum space $(k^0,\vec{k})$.  This is related to the
familiar Ricci tensor by the following steps. First, Fourier
transform the Ricci tensor of (\ref{einstein}) from coordinate space
to momentum space:
\bea
   R_{MN}(k) &=& \int dx^0\int d^{25} x~e^{ik\cdot x} R_{MN}(x) \nonumber \\
               {} &=& \half \{k^2 \epsilon_{MN} (k) - k_P k_M \epsilon^P_{\ N}
                - k_P k_N \epsilon^P_{\ M}
                + k_M k_N \epsilon^P_{\ P} \} \ .
\eea
Then, do the inverse Fourier transformation, but now with the
chemical potential $\mu$ instead of time $x^0$. In particular, this
defines
\bea
   R_{\mu\mu} (\mu ,\vec{x}) \equiv \int \frac{dk^0}{2\pi} \int d^{25}\vec{k}~e^{-ik^0\mu+i\vec{k}\cdot \vec{x}}
   R_{00}(k^0,\vec{k}) \ .
\eea

In our example, the time-time component of the Einstein equation
(\ref{einstein00}) is converted to an equation for thermal expectation
values
\be
  R_{\mu\mu} + 2\partial^2_{\mu}\phi = 8\pi T_{\mu\mu} \ ,
\ee
where $\mu = \ln z/\beta$ is the chemical potential, and the
stress tensor component $T_{\mu\mu}$ is the thermal expectation
value (\ref{thermexp}) in the limit $k^0\rightarrow 0$.
This particular component of the stress
tensor turns out to be a constant, the conserved initial energy.
Non-trivial chemical potential dependence would be obtained for what
where initially the pressure components $T_{ii}$ of the stress
tensor.

We have sketched how the thermal ensemble counterparts of the field
equations emerge from requiring scale invariance of a Dyson gas with
additional test charge density included.  In this way we could also
derive equations of motion of other spacetime fields, such as the
dilaton equation which we discussed in section 3.3.

\section{Discussion}

In this paper we  reformulated the worldsheet description of brane decay in terms of the statistical
mechanics of the Dyson gas.     The progress of time was marked by an increase in the average number
of particles in the gas.    This gave rise to a thermodynamic arrow of time -- the decrease of free
energy (or, equivalently, the increase of the Shannon entropy of the particle number distribution)
marked the passage of time.     The spacetime equations of motion translated into relations between the
expectation values of different statistical moments at different values of the chemical potential.
Our analysis has essentially involved rewriting worldsheet perturbation theory in a statistical
language\footnote{The electrostatics interpretation of worldsheet Green functions as
2d Coulomb potentials between
point charges is straightforward. The main issue
was that here it specifically leads to a statistical mechanical system with
a reinterpretation for time in target space.}, and we are not precisely describing the ``emergence'' of time from a timeless system.
Nevertheless, there are several lessons to be learned here concerning what it might mean for time to be
emergent.

First, in any scenario where time is emergent, the absence of time in the underlying microscopic
description will  require the latter to be some kind of Euclidean or statistical system.
In analogy with the AdS/CFT correspondence one might have expected some kind of relation between larger
scales in the statistical system and  later instants of time.
In the system we have studied, the ``scale'' in question turned out to be the number of interacting
particles in the system, and time as a continuous variable arose from this discrete quantity via
a Legendre transform relating it to a chemical potential.

Second, in a scenario where time emerges from a statistical system, one might have expected
a non-equilibrium flow to realize the flow in time.  In fact, this is a somewhat misleading expectation.
In non-equilibrium statistical mechanics, there is already a physical time and the dynamical equations
describe the flow {\it in time} of statistical moments towards their equilibrium values.
In our setup, we instead found that each instant of time was described by a different point of
equilibrium in a statistical system.  The spacetime equations of motion became relations between
these different points of equilibrium.  Indeed, it is possible that any emergent or holographic
description of time will have such a character.   We are attached to the idea of a ``flow in time''
partly because it is conventional to construct the initial value problem in terms of data
on a {\it spacelike} surface.  But we could equally well have formulated it on a {\it timelike} surface
of co-dimension one.   Indeed, the latter is the natural way to do things in AdS space.

Most of this paper was concerned with a statistical formulation of the rolling tachyon background of
Sen (\ref{boundarydef}).   However, in order to derive the equations of motion we required Weyl
invariance of deformations of the Dyson gas.   These deformations, involving the addition of external
charges, gave rise to new spacetime backgrounds solving the equations of motion.   Thus the set of
Weyl invariant deformations of the Dyson gas with $\beta = 2$ all give statistical descriptions of
spacetimes with decaying branes and different initial conditions\footnote{Note that although we considered
deformations at the boundary of the worldsheet corresponding to the one-dimensional Dyson gas,
this is just a specific example of a more general relation between scale invariant statistical mechanical
systems and (time-dependent) string backgrounds. For example, it would be interesting
to find a generalization for closed string tachyon condensation.}.   This is reminiscent of the
AdS/CFT correspondence where deformations of the CFT are dual to deformations of AdS spacetime.
{}From this perspective, one difference is that in AdS/CFT the deformations need not be
marginal -- relevant
deformations are dual to asymptotically AdS spacetimes that differ in the deep interior, and the RG flow
of the field theory is related to the spacetime equations of motion.

A simple way of exploring the role of non-marginal deformations of the Dyson gas is to consider
generalizations  of Sen's background (\ref{boundarydef}) that are of the form
\be
 \delta S_{open} = \lambda \int dt~e^{\alpha X^0(t)} \  .
\ee The parameter $\alpha$ in the exponent is related to the
inverse temperature of the associated Dyson gas $\beta$ by
$\alpha =\sqrt{\beta/2}$. For $\beta < 2$, the deformation is
relevant, for $\beta >2$ it is irrelevant.  This general system
has been studied in the random matrix literature, and is known as
the circular $\beta$-ensemble\footnote{Incidentally, a
longstanding problem was to find a random matrix model formulation
of the general $\beta$-ensembles. For circular ensembles, a
solution in terms of matrix models of certain sparse unitary
matrices was found in \cite{killipnenciu}.}. Dyson conjectured
that the canonical partition function of the ensemble at fixed $N$
is \be\label{dyson}
 Z_{N}(\beta) = \frac{\Gamma(1+\frac{\beta N}{2})}{[\Gamma(1+\frac{\beta}{2})]^{N}} \ .
\ee and various proofs have been presented in the literature (see
\cite{mehta}). Using Dyson's formula, we can construct the grand
canonical partition function \be\label{gc}
 Z_G(z,\beta,V) = \sum^\infty_{N=0} z^N_{\beta} Z_N(\beta) /N! \ ,
\ee with $Z_N(\beta)$ given by (\ref{dyson}).  It is readily shown
that the partition function diverges for $\beta > 2$ and converges
for $\beta < 2$, for all values of fugacity $z$. Exactly at $\beta=
2$, the sum is convergent  for $|z| < 1$, and requires analytic
continuation to larger values of $z$ as described before. This
suggests that $\beta = 2$ is a critical point for the Dyson gas (see
\cite{Fendley:1994ms,Fendley:1994rh}). Furthermore, the
renormalization group flow \cite{kanefisher} of the relevant
deformations ($\beta < 2$, convergent partition function) could
perhaps be related to the spacetime equations of motion as in
AdS/CFT and the worldsheet analyses of \cite{Freedman:2005wx}.

\bigskip

\noindent
{\bf \large Acknowledgments}

\bigskip

We thank Oren Bergman, Jan de Boer, Eric Gimon, Kurt Johansson, Per
Kraus, Antti Kupiainen, Finn Larsen, Albion Lawrence, Rob Leigh,
Subir Mukhopadhyay, Asad Naqvi, Sean Nowling, Amanda Peet, Koushik
Ray, Ashoke Sen, Diptiman Sen, David Tong, and Jung-Tay Yee for
useful comments and discussions. V.B. was supported in part by DOE
grant DE-FG02-95ER40893. E.K-V. and J.M. have been in part supported
by the Academy of Finland, by grant number 1210349. N.J. has been in
part supported by the Magnus Ehrnrooth foundation. This work was
also partially supported by the EU 6th Framework Marie Curie
Research and Training network ``UniverseNet'' (MRTN-CT-2006-035863).
N.J. and  J.M. thank the University of Pennsylvania, N.J. also
thanks the University of Illinois at Urbana-Champaign, and  E.K-V.
thanks the Aspen Center for Physics for hospitality while this work
was in progress.

\appendix
\section{Toward deriving equations of motion using Weyl variation method}

\subsection{Goal}\label{s1}

We would like to show that the equation of motions of the bulk massless fields, like graviton, dilaton etc. in presence of a decaying brane can be derived using Weyl invariance of the worldsheet non-linear sigma model. In absence of any boundary perturbation, the derivation is easy and given in Polchinski's book \cite{Polchinski:1998rq}. In presence of a boundary perturbation, the derivation is not so straightforward.

The sigma model is described on a curved worldsheet in presence of
massless closed string fields background and boundary perturbation
corresponding to decaying brane. In particular, for the
graviton-dilaton we would like to show that its equation of motion
is given by
\be\label{e1}
R_{MN}^L  +  2\nabla_{M}^L\nabla_{N}^L\phi = T_{MN}\,,
\ee
where $R^L_{MN}$ is
the linearized (in $h_{MN}$) target spacetime  Ricci tensor and
$T_{MN}$ is the graviton-dilaton one-point function in presence of
boundary perturbation , $\delta S_{\rm{bdry}} = \lambda\int ds
e^{X^0(s)}$ : \be\label{e2} T^M_{\ M} = \langle \CV
e^{-\lambda\int ds e^{X^0(s)}}\rangle_{\eta} \ , \ee where $\CV$
is the graviton-dilaton vertex operator defined in (\ref{vk}) and
$\langle\CO\rangle_{\eta} \equiv \int\CD X\CD h e^{-S_P}\CO$
implies that the correlation function is evaluated in the free
theory using perturbative methods. Here $S_P$ is the Polyakov
action defined in the first term of \refb{action}.

\bigskip

\subsection{Steps}\label{s2}

How can we derive (\ref{e1})? By enforcing Weyl invariance in the full quantum theory of the worldsheet
non-linear sigma model, we can achieve this. In this method we shall be able to derive
the analog of \refb{e1} in momentum space. We give the steps to be carried out in this section :
\begin{enumerate}

\item The starting point is \refb{taylorexp}. We write $Z_{\rm{disk}}[\varepsilon,\phi] =\CZ_0 +\CZ_1
+\ldots$, where $\CZ_n$ is the term with $n$-th power of $\CV$ :
$\CZ_n = \frac{g_c^n}{n!}\langle e^{-\delta S_{\rm{bdry}}} (:\CV:)^n\rangle_{\eta}$.
Since $\delta S_{\rm{bdry}}$ involves exactly marginal boundary perturbation, $\CZ_0$ is
invariant under worldsheet Weyl transformation : $\delta_W h_{ab} = \delta \omega(\sigma) h_{ab}$.
However, other terms $\CZ_n$ ($n>0$) are not. If we make sure that $\CZ_1$ is Weyl invariant,
all other terms with higher powers of $\CV$ will automatically be so. The Weyl variation of $\CZ_1$
is (the Polyakov action is Weyl invariant):
\bea\label{e3}
\delta_W\CZ_1 & = & \int\CD X \CD h\ e^{- S_P}\delta_W :\CV: \nonumber\\
              &   & +\int\CD X \CD h\ e^{- S_P} \sum_{N=1}^{\infty}\delta_W
              \left(\frac{(-\lambda\oint ds : e^{X^0(s)} :)^N}{N!} :\CV:\right) \ .
\eea
Weyl invariance is imposed by demanding $\delta_W\CZ_1 = 0$. The notation $:\quad :$ around an
operator implies that they are normal ordered.

\item The first term in the r.h.s. of the above equation can be derived by using the result given
in Polchinski's book. On curved worldsheet, first, we define normalized bulk operator like $\CV$ as:
\be\label{normalorder}
:\CV: = e^{{\rm D}_{{\rm bulk}}}\CV\,,\qquad
{\rm D}_{{\rm bulk}} = \int d^2\sigma_1 d^2\sigma_2
\Delta^{{\rm bulk}}_h(\sigma_1,\sigma_2)\frac{\delta}{\delta X^M(\sigma_1)}
\frac{\delta}{\delta X_M(\sigma_2)} \ .
\ee
Here $\Delta^{{\rm bulk}}_h(\sigma_1,\sigma_2)$ is the bulk-bulk propagator on curved worldsheet.
Note that in the flat worldsheet limit (+ boundary conditions)
\be\label{bulkprop}
\Delta^{{\rm bulk}}_h(\sigma_1,\sigma_2) \to \Delta^{{\rm bulk}}_{{\rm flat}}(z,\zbar;w,\bar{w})
= \half \ln | z - w|^2 \ .
\ee
Next we define the Weyl variation of the normalized bulk operator as
\be\label{weylbulk}
\delta_W:\CV: =  e^{\delta_W{\rm D}_{{\rm bulk}}}\CV + \delta_W^{{\rm exp}}\CV\,,
\ee
where the last term in the above equation denotes the explicit Weyl variation.
The Weyl variation of $\D_{{\rm bulk}}$ is
\be
\delta_W{\rm D}_{{\rm bulk}} = \int d^2\sigma_1 d^2\sigma_2
\delta_W\Delta^{{\rm bulk}}_h(\sigma_1,\sigma_2)\frac{\delta}{\delta X^M(\sigma_1)}
\frac{\delta}{\delta X_M(\sigma_2)} \ .
\ee

\item Choosing a particular RG scheme\footnote{We put $\gamma=0$ in eqns (3.6.17a-c)
of \cite{Polchinski:1998rq}.} on the worldsheet, it turns out that
\be\label{weylvk}
\delta_W:\CV: = g_c\int d^2\sigma\sqrt h \delta\omega\Big[\widetilde{R}'^L_{MN}(k)h^{ab}
:\partial_a X^M\partial_b X^N e^{ik\cdot X}:
+ \alpha' F\widetilde{R} : e^{ik\cdot X}:\Big]\,,
\ee
where
\bea\label{e4}
\widetilde{R}'^L_{MN}(k) &=& \half \{-k^2\varepsilon_{MN} + k^{N}k^{P}\varepsilon_{MP}
+ k^{M}k^{P}s_{NP} - k_{M}k_{N} \varepsilon^{P}_{\ P} +4k_Mk_N\phi\}\nonumber\\
\mbox{} &=& \widetilde{R}^L_{MN}(k) + 2k_Mk_N\phi \nonumber \\
F & = & -\half k^2\phi \ .
\eea

\item The most non-trivial part for evaluating the r.h.s. of (\ref{e3}) is the second term.

We need to generalize the rule in \refb{weylbulk} to the case when we have a composite operator
made of bulk and boundary operators on the curved worldsheet. We would like to give a recipe for
this generalized definition of {\em normal ordering} next.

\item {\em Polchinski's Rule Generalized} :
We use $\sigma_i= (\sigma_i^0, \sigma_i^1,)$ for bulk and $(t_i, u_j)$ for boundary coordinates.
We define the following two operators $\D_{{\rm bdry}}$ and $\D_{{\rm bulk-bdry}}$ on the boundary
and bulk-boundary respectively,
\bea\label{e12}
\D_{{\rm bdry}} &\equiv& \half\int du_1\ du_2\Delta_{{\rm bdry}}(u_1,u_2)
\frac{\delta}{\delta X^{\rho}(u_1)}
\frac{\delta}{\delta X_{\rho}(u_2)}\nonumber\\
\D_{{\rm bulk-bdry}} &\equiv& \half\int d^2\sigma_3\ du_3\Delta_{\rm{bulk-bdry}}(\sigma_3,u_3)
\frac{\delta}{\delta X^{\rho}(\sigma_3)}
\frac{\delta}{\delta X_{\rho}(u_3)} \ .
\eea
In our case the normal ordered boundary operator is $:e^{X^0(s)}: = e^{\D_{{\rm bdry}}} e^{X^0(s)}$
which is exactly marginal. Hence its explicit Weyl variation as well as the Weyl variation
of $\D_{{\rm bdry}}$ are vanishing.
Keeping these two points in mind, we define the following generalized differential operator which
does not include $\D_{{\rm bdry}}$:
\be\label{e13}
\D \equiv \D_{{\rm bulk}} + \D_{{\rm bulk-bdry}} \ .
\ee
Let $\CF$ be a {\em composite} operator made of bulk and boundary operators. In our case $\CF$ will
be $\prod_i (-\lambda)e^{X^0(s_i)}\CV$. We define the Weyl variation of the normal ordered $\CF$ as
\be\label{e15}
\delta_W:\CF: = (\D\delta_W\D)_{\ast}\CF \ .
\ee
The subscript ${}_{\ast}$ denotes an important rule : we set terms like
$(\D_{\rm{bulk-bdry}})^n = 0$ for $n > 1$. The reason is that there is no concept of self-contraction for computing correlators between bulk and boundary operators. The explicit Weyl variation of $\CF$ in our case is zero since the variation for both $\prod_i e^{X^0(s_i)}$ and $\CV$ are zero.
Equation (\ref{e15}) will be our working definition.

\end{enumerate}

\bigskip

\subsubsection{Explicit calculations}\label{ss33}

From (\ref{e12}) and (\ref{e13}), we find that
\be\label{e16}
(\D\delta_W\D)_{\ast}\CF = \left(\D_{\rm{bulk}}\delta_W\D_{\rm{bulk}}
+  \D_{\rm{bulk}}\delta_W\D_{\rm{bulk-bdry}} + \D_{\rm{bulk-bdry}}\delta_W\D_{\rm{bulk}} \right)\CF \ .
\ee
In our case
\be\label{e17}
\CF = \int d^2\sigma\sqrt h \Big[\varepsilon_{MN}h^{ab}\partial_aX^{M}\partial_bX^{N}e^{ikX}(\sigma)
+ \alpha'\widetilde{R}\phi e^{ik\cdot X}\Big]\int\prod_k ds_k\prod_j (-\lambda)e^{X^0(s_j)}
\ee
For $\CF$ given in \refb{e17}, we compute its Weyl variation $\delta_W :\CF:$, following \refb{e15}.
So we need to evaluate (\ref{e16}). The result is :
\bea\label{e18}
& &(\D\delta_W\D)_{\ast}\ \CF\nonumber\\
&=& \int d^2\sigma\prod_k ds_k \sqrt h h^{ab}\varepsilon_{MN}\Bigg[-
4k^{M}k^{N} \left(\partial_a\delta_W\Delta_{{\rm
bulk}}\partial_b\Delta_{{\rm bulk}}
+ \partial_a\Delta_{{\rm bulk}}\partial_b\delta_W\Delta_{{\rm bulk}}\right)\nonumber\\
&-& 2k^2\Big\{\eta^{MN}(\partial_a\partial_b'\delta_W\Delta_{{\rm bulk}})\Delta_{{\rm bulk}} +
\eta^{MN}(\partial_a\partial_b'\Delta_{{\rm bulk}})\delta_W\Delta_{{\rm bulk}} \nonumber\\
&+&  ik^{(M}\Big(\partial_{(a}\delta_W\Delta_{{\rm bulk}}\Big)\Delta_{{\rm bulk}}\partial_{b)} X^{N)}\Big\}
\nonumber\\
&+& 2i\eta^{0(M}k^{N)}\sum_i\partial_{(a}\delta_W\Delta_{{\rm bulk-bdry}}(\sigma,s_i)
\partial_{b)}\Delta_{{\rm bulk}} \nonumber\\
&+& 2ik^0\eta^{MN}\sum_i\delta_W\Delta_{{\rm bulk-bdry}}(\sigma,s_i)
\partial_a\partial_b'\Delta_{{\rm bulk}}  \nonumber\\
&+& 2ik^0\eta^{MN}\partial_a\partial_b\delta_W\Delta_{{\rm bulk}}
\sum_i \Delta_{{\rm bulk}}(\sigma,s_i) \nonumber\\
&+& 2ik^{(M}\eta^{0N)} \partial_{(a}\delta_W\Delta_{{\rm bulk}} \sum_i \partial_{b)}
\Delta_{{\rm bulk-bdry}}(\sigma,s_i)\Bigg]e^{ikX} \prod_j (-\lambda)e^{X^0(s_j)} \ .
\eea
This equation actually evaluates Weyl variation of $\langle \CV e^{-\delta S_{\rm{bdry}}}\rangle_h$
where the subscript $h$ denotes curved worldsheet. This is easy to verify : first, we compute one-point
function of $\CV$ in presence of boundary perturbation on flat worldsheet,
$\langle \CV e^{-\delta S_{\rm{bdry}}}\rangle_{\eta}$. Next, we covariantize this result
for curved worldsheet to obtain $\langle \CV e^{-\delta S_{\rm{bdry}}}\rangle_h$.
Finally we can compute its Weyl variation by using the rules in \refb{weylbulk}, \refb{e13} and \refb{e15}.

The next step is to express the r.h.s. of \refb{e18} as $\sim \delta\omega T^M_{\ M}$.
Finally we have to substitute \refb{e18}, \refb{weylvk} and \refb{e4} in r.h.s. of \refb{e3}
to obtain the momentum space version of  equation of motion in \refb{einstein}.
We leave this issue for future work.

\end{document}